\documentclass{raa}            

\usepackage{graphicx,times}             
\usepackage{natbib}
\usepackage{amssymb,amsmath}
\bibpunct{(}{)}{;}{a}{}{,}

\usepackage[pagebackref=true]{hyperref}

\usepackage{float}

\usepackage{indentfirst}

\newcounter{o}
\newcounter{t}
\newcounter{f}
\graphicspath{{./}}

\begin{document}

  \title{Radiative MHD Studies of Solar Spicules: Driving Mechanisms and the Role of Background Magnetic Field
}

   \volnopage{Vol.0 (202x) No.0, 000--000}      
   \setcounter{page}{1}          

   \author{Jin Li 
      \inst{1,2}
   \and Lei Ni
      \inst{1,2,3,*}\footnotetext{$*$Corresponding Authors, these authors contributed equally to this work.}
   \and Robertus Erd\'{e}lyi
      \inst{4,5,6}
   \and Jun Lin
      \inst{1,2,3}
   \and Xiaoli Yan
      \inst{1,2,3}
   \and Guanchong Cheng
      \inst{1,2,3}
   }

   \institute{Yunnan Observatories, Chinese Academy of Sciences,
             Kunming, Yunnan 650216, China; {\it leini@ynao.ac.cn}\\
        \and{University of Chinese Academy of Sciences, Beijing 100049, China}\\
        \and{Yunnan Key Laboratory of Solar Physics and Space Science, Kunming 650216, China}\\
        \and{Solar Physics and Space Plasma Research Centre, School of Mathematical and Physical Sciences, 
            University of Sheffield, Hicks Building, Hounsfield Road, Sheffield, S3 7RH, UK}\\
        \and{Department of Astronomy, E\"{o}tv\"{o}s Lor\'{a}nd University, P\'{a}zm\'{a}ny P\'{e}ter s\'{e}t\'{a}ny 1/A, 
            Budapest, H-1112, Hungary}\\
        \and{Gyula Bay Zolt\'{a}n Solar Observatory (GSO), Hungarian Solar Physics Foundation (HSPF), 
            Pet\"{o}fi t\'{e}r 3, Gyula, H-5700, Hungary}\\
\vs\no
   {\small Received 202x month day; accepted 202x month day}}

\abstract {We conduct 2.5D radiative magnetohydrodynamic (MHD) simulations to investigate the driving mechanisms of the solar spicules in coronal holes and how the different background magnetic fields affect their formation. The simulation model includes the upper convection zone, photosphere, chromosphere, and lower corona. We run several cases with different background magnetic fields to primarily explore the effects of magnetic field strength and inclination on the characteristics of the formed spicules, such as the maximum height, lifetime, maximum upward velocity, and deceleration. According to the results, we find that a weaker background magnetic field tends to cause solar spicules that exhibit higher heights, longer lifetimes and faster maximum upward velocities. Additionally, when the background magnetic field inclines, the generated spicules incline accordingly; compared with those in a vertical magnetic field, the spicules in the inclined magnetic field are relatively longer, with smaller decelerations, longer lifetimes and lower maximum upward velocities. Meanwhile, by tracking and analyzing the formation processes of the spicules in two cases with different magnetic field strengths, we find that most of the spicules are mainly driven by shock waves induced by convective and turbulent motions around the solar surface, while less than one third of the spicules are primarily driven by high-velocity reconnection outflows. These results provide a more in-depth basis for the theoretical understanding of the driving mechanisms and formation processes of solar spicules.
\keywords{Sun: activity --- Sun: photosphere --- Sun: chromosphere
}
}

   \authorrunning{J. Li et al. }            
   \titlerunning{Radiative MHD Studies of Solar Spicules}  
   \maketitle

\section{Introduction}           
\label{sect:intro}

Solar spicules were first reported by \citet{Secchi1877Astronomia}. They are highly active, dynamic jet structures commonly observed in the solar chromosphere and are distributed almost across the entire solar surface. The lifetime of the spicules is typically less than $10$\,minutes \citep{Pereira2012Quantifying}. The most popular bands for spicule observations include H$\alpha$, Ca \Roman{t}, Mg \Roman{t}, Si \Roman{f} and He \Roman{t}. When spicules are observed at the solar limb, they often appear as dense and elongated structures (e.g., \citealp{Sterling2000SoPh}). Some scholars argue that there are two types of spicules \citep{DePontieu2007TaleTwoSpicules,Pereira2012Quantifying}. 

Based on the observational statistics from the Solar Optical Telescope (SOT, \citealp{Tsuneta2008SoPh,Suematsu2008SoPh}) in the Ca \Roman{t} H band, type \Roman{o} spicules have a maximum height in the range of $4-8$\,Mm, a maximum upward velocity of approximately $15-40$\,km\,s$^{-1}$, and a lifetime of about $150-400$\,s, with their distribution primarily in active regions. In contrast, type \Roman{t} spicules exhibit a maximum height concentrated between $3-9$\,Mm, a maximum upward velocity of roughly $30-110$\,km\,s$^{-1}$, and a lifetime of around $50-150$\,seconds, and they dominate in quiet regions and coronal holes, where the background magnetic field is relatively weaker \citep{Pereira2012Quantifying}. Moreover, type \Roman{o} spicules exhibit two phases when observed in the chromospheric bands: an upward phase and a downward phase. In contrast, type \Roman{t} spicules typically only show upward motions in these same bands. 

Later, observations in high-temperature bands such as those from the Interface Region Imaging Spectrograph (IRIS, \citealp{DePontieu2014}) and the Atmospheric Imaging Assembly (AIA, \citealp{Lemen2012AIA}) instruments on board the Solar Dynamics Observatory (SDO, \citealp{Pesnell2012SDO}) have revealed that some type \Roman{t} spicules are heated to temperatures above the transition region. Furthermore, the upward and downward phases are both observed in these high-temperature bands respectively, and the lifetime of these type \Roman{t} spicules can reach $500 - 800$\,s \citep{Pereira2014ApJL}. However, there is still controversy over whether Type \Roman{t} spicules constitute a new type of spicule (e.g., \citealp{Zhang2012Revision}). Spicules often have fine internal structures and exhibit a multi-threaded characteristic. These threads can split and merge over time \citep{Skogsrud2014Multithreaded}. Among them, some threads can reach transition region temperatures, while others remain at chromospheric temperatures \citep{Pereira2014ApJL,Skogsrud2015ApJ,Rouppe2015ApJL,Dover2020,Dover2022}. The waves and oscillations in spicules have long been studied (e.g., \citealp{Zaqarashvili2009OscillationsWaves}). In addition, \citet{Qi2022A&A} found that higher and faster H$\alpha$ jets are more likely to elicit coronal responses.

The driving mechanisms of spicules are not unique, but they are generally related to convective and turbulent motions around the solar surface, which can lead to the formation of magnetohydrodynamic waves, the complex magnetic field structures for generating upward Lorentz force, and the upward pressure gradient. The generated slow mode shock waves in the lower atmosphere are believed as one of the driving mechanisms \citep{Parker1964ApJ,Hollweg1982Origin,DePontieu2004Nature,Hansteen2006DynamicFibrils}. The shearing and rotational motions of magnetic flux tubes can trigger the generation of Alfv\'{e}n waves and create an upward Lorentz force, which drives plasma upward to form spicules \citep{Iijima2017ApJ,Liu2019Evidence,Oxley2020ApJ,Scalisi2021ApJ}. Magnetic reconnection between magnetic fields with opposite directions is also one of the possible mechanisms to cause spicule formation \citep{Yokoyama1995Nature,Archontis2010A&A,Singh2011PhPl,Samanta2019Science}. In the low-temperature, partially ionized lower atmosphere, the decoupling of ions and neutral (ambipolar diffusion) can amplify the Lorentz force, which triggers the formation of longer and faster type \Roman{t} spicules \citep{DePontieu1998A&A,martinez-sykora2017generation}.

Though the background magnetic fields in the lower corona are generally vertical to the solar surface in regions such as the center of coronal holes (e.g., \citealp{Tu2005,Yang2020Global}), the magnetic inclination angle is not zero in most regions above the solar surface. Therefore, it is necessary to investigate how inclined magnetic field affects the generated spicules (e.g., \citealp{DePontieu2004Nature}). \citet{Heggland2011ApJ} conducted 2D numerical simulations using the Bifrost code and found that the jets formed in inclined magnetic fields have longer lifetimes and smaller decelerations. In addition, the background magnetic field strength distribution on the solar surface is not uniform, and it is also necessary to explore the influence of magnetic field strength on the generated spicules. \citet{Kesri2024ApJ} performed 2D simulations using the PENCIL code and observed that as the magnetic field strength increases, the maximum height of the spicules decreases accordingly.   

In this paper, the driving mechanisms of solar spicules were investigated through 2.5D MHD simulations, under a relatively realistic plasma environment that includes radiative cooling, gravitational stratification from the upper convective zone to the lower coronal, and dramatic temperature variations. In addition, we conducted a comparative analysis of the variations in spicule characteristics, such as the evolutionary process, lifetime, deceleration, height, and maximum upward velocity, under environments with different background magnetic fields. In Section 2, we introduce the numerical models and methods applied in this work. The numerical results are presented in Section 3. We give the summary and provide a critical discussion in Section 4.

\section{Models and methods}

\subsection{MHD model}

We used the developed NIRVANA code \citep{Ziegler2008NIRVANA,Ni2021A&A,Cheng2024ApJL,Wang2025ApJ} to perform 2.5D MHD simulations. The upper convection zone, photosphere, chromosphere, and lower corona were included in the simulations. The format of the solved MHD equations are as follows:
\begin{equation}
\frac{\partial\rho}{\partial t}=-\nabla\cdot(\rho\boldsymbol{v}),
\end{equation}
\begin{equation}
\frac{\partial(\rho \boldsymbol{v})}{\partial t}=-\nabla\cdot[\rho \boldsymbol{vv}+(P+\frac{1}{2\mu_0}|\boldsymbol{B}|^{2})I-\frac{1}{\mu_0}\boldsymbol{BB}]+\rho \boldsymbol{g},
\end{equation}
\begin{equation}
\frac{\partial e}{\partial t}=-\nabla\cdot[(e+P+\frac{1}{2\mu_0}|\boldsymbol{B}|^{2})\boldsymbol{v}]+\nabla\cdot[\frac{1}{\mu_0}(\boldsymbol{v}\cdot\boldsymbol{B})\boldsymbol{B}]
+\nabla\cdot[\frac{\eta}{\mu_0}\boldsymbol{B}\times(\nabla\times\boldsymbol{B})]-\nabla\cdot\boldsymbol{F_\mathrm{c}}
+\rho\boldsymbol{g}\cdot\boldsymbol{v}+Q_{\mathrm{rad}},
\end{equation}
\begin{equation}
\frac{\partial\boldsymbol{B}}{\partial t}=\nabla\times(\boldsymbol{v}\times\boldsymbol{B}-\eta\nabla\times\boldsymbol{B}),
\end{equation}
\begin{equation}
e=\frac{P}{\gamma-1}+\frac{1}{2}\rho|\boldsymbol{v}|^2+\frac{1}{2\mu_0}|\boldsymbol{B}|^{2},
\end{equation}
\begin{equation}
P=\frac{(1.1+Y_{\mathrm{iH}}+0.1Y_{\mathrm{iHe}})\rho}{1.4m_\mathrm{i}}k_\mathrm BT,
\end{equation}
where $\rho$ is the plasma mass density, $\boldsymbol{v}$ is the fluid velocity,  $\boldsymbol{B}$ is the magnetic field, $\mu_0=4\pi\times10^{-7}\ \mathrm N\  \mathrm A^{-2}$, is the vacuum magnetic permeability, $I$ is the unit tensor, $\boldsymbol{g}=273.93\ \mathrm m\ \mathrm s^{-2}$, is the gravitational acceleration of the Sun, $e$ is the energy density, $P$ is the plasma pressure, $\gamma=5/3$, is the ratio of specific heats, $Y_{\mathrm{iH}}$ and $Y_{\mathrm{iHe}}$ are ionization fractions of hydrogen and helium, respectively, $m_{\mathrm i}$ is the mass of a hydrogen ion, $k_\mathrm B=1.3806\times10^{-23}\ \mathrm J\ \mathrm K^{-1}$, is the Boltzmann constant, $T$ is the temperature, $Q_\mathrm{rad}$ refers to the radiative cooling, and $\nabla\cdot F_\mathrm c $ refers to the heat conduction. The explicit viscosity terms and ambipolar diffusion terms are not included in this work.

In the upper convection zone and photosphere, assuming the plasma is in a state of local thermodynamic equilibrium, the ionization degree of hydrogen $Y_{\mathrm{iH}}$ satisfies the Saha equation and it depends on the temperature and density of the plasma \citep{Cheng2024MagneticReconnection}. In the chromosphere, $Y_{\mathrm{iH}}$ is derived from the temperature-dependent tables supplied by \citet{Carlsson2012Approximations}. In the corona, the plasma is assumed to be fully ionized. In the simulations, the number density of helium is set to $10$\% of the total number density of hydrogen. A simple temperature-dependent ionization degree of helium $Y_{\mathrm{iHe}}$ is deduced, according to the experiences by solving the radiative transfer equations \citep{Carlsson2012Approximations,Ni2022A&A}.

Further, $\eta$ is the magnetic diffusion coefficient, and is given by:
\begin{equation}
\eta=\eta_{\mathrm{ei}}+\eta_{\mathrm{en}}=\frac{m_\mathrm e\nu_\mathrm{ei}}{e_{\mathrm c}^2n_{\mathrm e}\mu_0}+\frac{m_{\mathrm e}\nu_{\mathrm{en}}}{e_{\mathrm c}^2n_{\mathrm e}\mu_0},
\end{equation}
where $m_\mathrm e$, $e_\mathrm c$, $n_\mathrm e$, $\nu_\mathrm {ei}$ and $\nu_\mathrm{en}$ are the mass of electron, the electron charge, the electron density, the electron-ion collision and the electron-neutral collision frequencies, respectively. The coefficients $\nu_\mathrm {ei}$ and $\nu_\mathrm{en}$ are given by:
\begin{equation}
\nu_{\mathrm{ei}}=\frac{n_\mathrm ee_{\mathrm c}^4\Lambda}{3m_\mathrm e^2\epsilon_{0}^2}(\frac{m_\mathrm e}{2\pi k_\mathrm BT})^{\frac{3}{2}},
\end{equation}
\begin{equation}
\nu_{\mathrm{en}}=n_\mathrm n\sqrt{\frac{8 k_\mathrm BT}{\pi m_{\mathrm{en}}}}\sigma_{\mathrm{en}},
\end{equation}
where $\epsilon_{0}$ is the permittivity of vacuum, $\sigma_{\mathrm{en}}$ is the collision cross-section, $n_\mathrm n$ is the number density of the neutral particle, $m_{\mathrm{en}}=m_\mathrm em_\mathrm n/(m_\mathrm e+m_\mathrm n)$, $m_\mathrm e$ and $m_\mathrm n$ are the mass of electron and neutral particle, respectively. Since $m_\mathrm n$ is much larger than $m_\mathrm e$, we can get $m_{\mathrm{en}}\cong m_\mathrm e$. $\Lambda$ is the Coulomb logarithm, and is given by:
\begin{equation}
\Lambda=23.4-1.15\mathrm{log}_{10}n_\mathrm e+3.45\mathrm{log}_{10}T,
\end{equation}
where $n_\mathrm e$ is expressed in the CGS unit system, and the unit of $T$ is eV. The electron-neutral collision is contributed by both the collisions between electrons and neutral hydrogen, and the collisions between electrons and neutral helium. 

\subsection{Radiation cooling models and heat conduction}

Four different radiation cooling models were applied in the upper convection zone, photosphere, chromosphere, and lower corona, respectively. The radiation cooling of each layer is denoted by $\mathrm{Q}_{rad1}$, $\mathrm{Q}_{rad2}$, $\mathrm{Q}_{rad3}$ and $\mathrm{Q}_{rad4}$ in sequence, the complete expression for radiative cooling is as follows (e.g., \citealp{Wang2025ApJ}):
\begin{equation}
Q_{\mathrm{rad}}=\begin{cases}Q_{\mathrm{rad1}},\ m_\mathrm c >m_{\mathrm{c1}}\\
Q_{\mathrm{rad2}},\ m_{\mathrm{c2}}<m_\mathrm c\leq m_{\mathrm{c1}}\\
Q_{\mathrm{rad3}},\ m_{\mathrm{c3}}<m_\mathrm c\leq m_{\mathrm{c2}}\\
Q_{\mathrm{rad4}},\ m_\mathrm c\leq m_{\mathrm{c3}},
\end{cases}
\end{equation}
where $m_\mathrm c$ is the neutral hydrogen column mass, $m_{\mathrm{c1}}=10^{1.56}\ \mathrm {kg\ m}^{-2}$, $m_{\mathrm{c2}}=10^{-1.6}\ \mathrm {kg\ m}^{-2}$, $m_{\mathrm{c3}}=10^{-4.7}\ \mathrm {kg\ m}^{-2}$. $m_\mathrm c$ is calculated by:
\begin{equation}
m_{\mathrm{c}}(x,y)=\int _y^\infty \rho(x,y)dy,
\end{equation}
where $\rho(x,y)$ is the mass density at grid$(x,y)$.

In the convection zone, we used the model proposed by \citet{Robinson2003MNRAS}, where the expression for the radiative cooling is as follows:
\begin{equation}
Q_{\mathrm{rad1}}=\nabla\cdot(k_{\mathrm{rad}}\nabla T).
\end{equation}
Here $k_{\mathrm{rad}}=16\sigma T^3/(3k_\mathrm R)$ is the coefficient of conduction for radiative transport \citep{Kippenhahn1994}, $\sigma=5.67\times 10^{-8}\ \mathrm J\ \mathrm s^{-1}\ \mathrm m^{-2}\ \mathrm K^{-4}$ is the Stefan-Boltzmann constant, and $k_\mathrm R$ is the atomic Rosseland mean opacity of the sun-like star model from the OPAL tables \citep{rogers1992radiative}.

In the photosphere, for the radiative cooling we employed the model proposed by \citet{Abbett2012SoPh}, with the expression as follows:
 \begin{equation}
Q_{\mathrm{rad2}}=-2k^\mathrm B\rho\sigma T^4E_2(\tau^\mathrm B),
\end{equation}
where $k^\mathrm B$ is the Planck-averaged opacity, $\tau^\mathrm B$ is the optical depth, and $E_2$ is a function depending on the $\tau^\mathrm B$. 

In the chromosphere, we used the approximate model proposed by \citet{Carlsson2012Approximations}, with the expression given below:
 \begin{equation}
Q_{\mathrm{rad3}}=-\sum_{\mathrm{X=H,Mg,Ca}}L_{\mathrm X_m}(T)E_{\mathrm X_m}(\tau)\frac{N_{\mathrm X_m}}{N_{\mathrm X}}(T)A_\mathrm X\frac{N_\mathrm H}{\rho}n_\mathrm e\rho,
\end{equation}
where $L_{\mathrm X_m}$ is the optically thin radiative loss function per electron an per particle of element X in ionization stage $m$, $E_{\mathrm X_m}(\tau)$ is the photon escape probability as function of the depth parameter $\tau$, which depends on the column mass, $\frac{N_{\mathrm X_m}}{N_\mathrm X}$ is the fraction of element X in ionization stage $m$, $A_\mathrm X$ is the abundance of element X. Here, X comprises H, Ca and Mg. According to the solar atmospheric model, $A_\mathrm H=1$, $A_{\mathrm{Mg}}=3.885\times10^{-5}$, $A_{\mathrm{Ca}}=2.042\times10^{-6}$.

In the corona, the radiative cooling is simplified as a radiative loss that depends only on temperature and density, and is given as follows:
 \begin{equation}
Q_{\mathrm{rad4}}=-n_\mathrm Hn_\mathrm ef(T),
\end{equation}
where $n_\mathrm H$ and $n_\mathrm e$ are the number densities of hydrogen and electrons, respectively, and $f(T)$ is a function of temperature \citep{Gudiksen2011}.

In addition, we included the heat conduction term along the magnetic field lines. The heat flux ${\boldsymbol F}_{\rm c}$ is given by:
\begin{equation}
{\boldsymbol F}_{\mathrm c}=-k_{\parallel}(\nabla T\cdot\boldsymbol{\hat {B}})\boldsymbol{\hat {B}},
\end{equation}
where $\boldsymbol{\hat {B}}= \boldsymbol B/|\boldsymbol B|$, is the unit vector in the direction of magnetic field. $k_{\parallel}$ is the parallel thermal conductivity coefficient in J\,K$^{-1}$\,m$^{-1}$\,s$^{-1}$, which is given by Spitzer theory \citep{spitzer1962}:
\begin{equation}
k_{\parallel_{\mathrm{sp}}}=\frac{1.84\times10^{-10}}{\Lambda}T^{\frac{5}{2}}.
\end{equation}

\subsection{Initial setups and boundary conditions}

In this paper, the simulation domain of each case is the same. The $y$-direction is perpendicular to the solar surface, ranging from $-3$ to $22$\,Mm (from the upper convection zone to the low corona), where the segment from $-3$ to $0$\,Mm corresponds to the convection zone. The $x$-direction is parallel to the solar surface, ranging from $-6$ to $6$\,Mm. The initial plasma parameters above the solar surface were set with reference to the C7 model \citep{Avrett2008ApJS}, while those below the solar surface were set with reference to the Standard Solar Model. Therefore, the initial plasma density varies with height in the $y$-direction, and they are uniform in the $x$-direction.

Periodic boundary conditions were applied in the $x$-direction. The boundary conditions for the bottom and the top boundary corresponded to "inflow" and "outflow" (see e.g., \citealp{Wang2025ApJ}), respectively. By implementing such boundary conditions in the $y$-direction, the fluid is allowed to flow into the domain at the bottom boundary but not out, while at the top boundary, it is allowed to flow out of the domain but not in. Additionally, we incorporated a strong artificial dissipation in the range of $20$ to $22$\,Mm along the $y$-axis at the top boundary. Such outflow boundary conditions and strong artificial dissipation can significantly reduce the wave reflection from the top boundary back into the simulation domain. Though the "inflow" boundary was applied at the bottom, there are no mass and energy flowing into the simulation domain through the bottom boundary. The mass and energy continually flow out the simulation domain from the top boundary, and the average mass and energy fluxes there are about $10^{-9}$\,kg\,m$^{-2}$\,s$^{-1}$ and $100$\,W\,m$^{-2}$ respectively, which are consistent with the values measured from the recent numerical simulation performed by using the advanced MURaM code \citep{Chen2025}. Such results indicate that the simulations are not influenced by external factors and the numerical results are reliable during the period shown in this paper.

A total of five cases were conducted in this work. Among them, Case A has relatively lower resolution, with $256\times1024$ uniform grid points applied in the $x$- and $y$-directions, respectively, while the other four cases feature higher resolution, utilizing $896\times2048$ grid points. The initial uniform magnetic fields were set to be perpendicular to the solar surface in Cases A, B, C and D, with magnetic field strengths of $20$\,G, $20$\,G, $10$\,G and $4$\,G, respectively. After Case D reached $t=2000$\,s, we introduced uniform magnetic fields of $16$\,G and $8$\,G in the x and y directions, respectively, to create a new simulation branch designated as Case E. Meanwhile, the original Case D continued running without modification. Eventually, the background magnetic field strength in Case E is about $20$\,G, and the magnetic inclination angle in the corona is about $53.13^{\circ}$. The grids and initial magnetic fields in the five cases are summarized in Table~\ref{tab1}.

\begin{table}
  \begin{center}
  \caption[] {The five cases with different grids and initial magnetic fields.}
  \label{tab1}
  \begin{tabular}{|l|l|l|}
  \hline
  Model & Grid Cells & Initial Magnetic Field ($B_0$)\\
  \hline
  Case A & $256\times1024$ & $20$\,G, vertical \\
  \hline
  Case B & $896\times2048$ & $20$\,G, vertical \\
  \hline
  Case C & $896\times2048$ & $10$\,G, vertical \\
  \hline
  Case D & $896\times2048$ & $4$\,G, vertical \\
  \hline
  Case E & $896\times2048$ & $20$\,G, inclined \\
  \hline
  \end{tabular}
 \end{center} 
\end{table}

\subsection{Generated convective motions}   

\begin{figure}[H]
   \centering
   \includegraphics[width=12cm, angle=0]{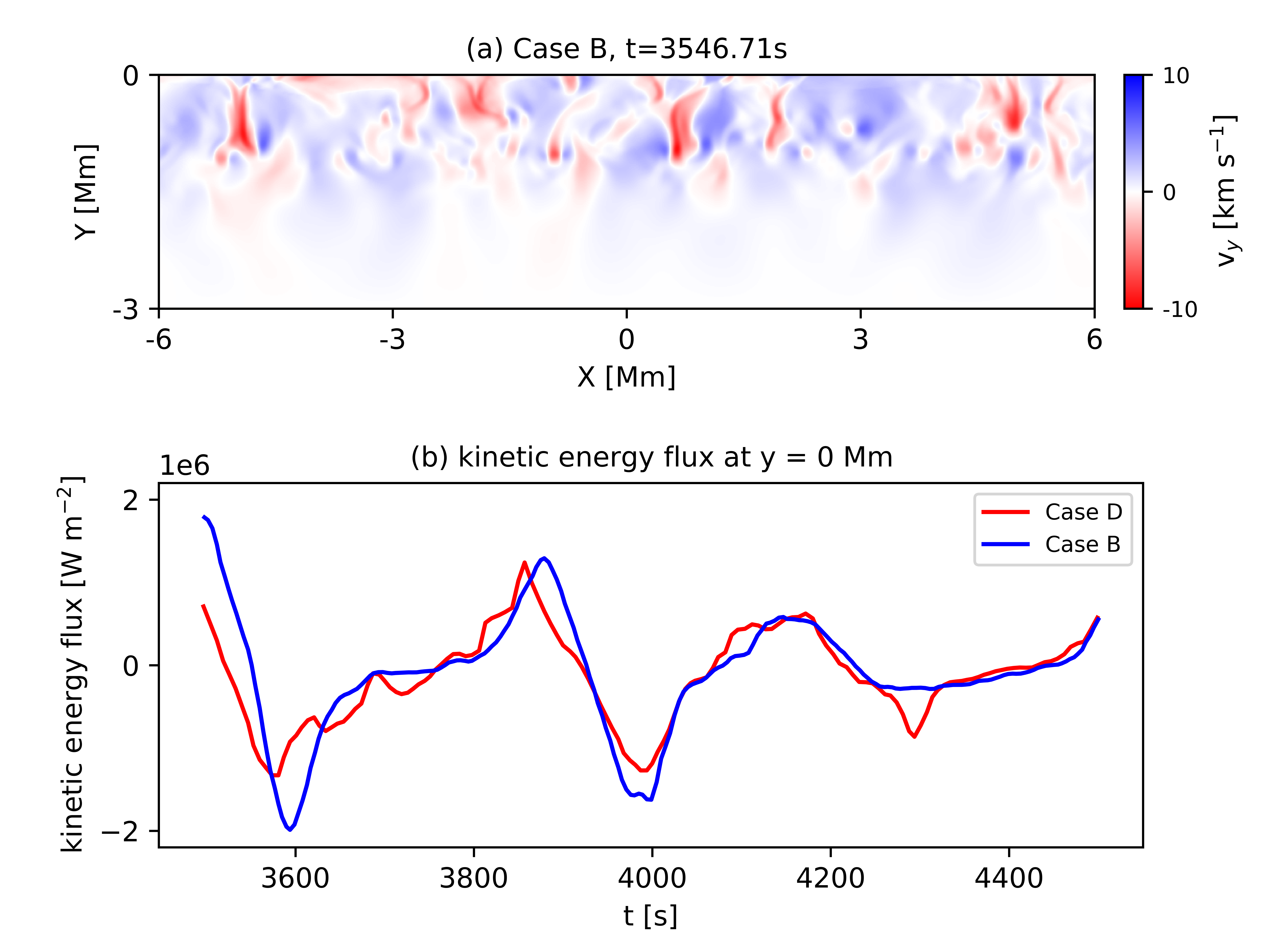}
   \caption{Distribution of the velocity in the $y$-direction in the convection zone in Case B at $t=3546.71$\,s is presented in panel (a). The time evolutions of the kinetic energy flux in the $y$-direction in Cases B and D are shown in panel (b). The kinetic energy flux is the averaged value along the $x$-direction.}
   \label{fig1}
   \end{figure}

In this work, the convective motions were self-consistently generated in the convection zone of the simulation box. Since the system was not in equilibrium at the beginning, it had to undergo a self-adjustment process in the early stage, and eventually reached a dynamic equilibrium state after about $t=2500$\,s. Figure~\ref{fig1}(a) shows the distributions of velocity in the $y$-direction ($v_y$) in the convection zone at $t=3546.71$\,s. The regions with red color represent the downflows and the regions with blue colors represent the upflows. One can find that the upflows are broader and the downflows are narrower. Between $y=0$ and $y=-1.5$\,Mm, downflows have typical speeds of $4-8$\,km\,s$^{-1}$, whereas upflows have typical speeds of only $2-4$\,km\,s$^{-1}$. The asymmetry between the upflows and the downflows is a robust feature of compressible convection in a stratified medium: ascending material expands and descending material is compressed. One can also find the similar descriptions of the convective velocities from the previous MURaM simulation in Section 3.3.1 of Prof. Mark Cheung's PhD dissertation in 2006 \citep{Cheung2006}. The kinetic energy flux at the solar surface (Figure~\ref{fig1}(b)) is also close to the values measured from the recent numerical simulations performed by using the advanced MURaM code (e.g., \citealp{Chen2025,Cheng2025}).

One can also note that the convective motions in the lower half of the convective zone are much weaker, which then results in much weaker perturbations on magnetic fields. The reason is that the boundary condition applied in this work causes the velocities gradually approaches zero near the bottom boundary. However, as we discussed above, the convective motions in the upper part of the convection zone and the kinetic energy flux at the solar surface are very close to the results from the MURaM code. Therefore, we can just simply ignore the results below $y=-1.5$\,Mm, which are far way from the regions we focus on. Since no mass and energy flow into the simulation domain to replenish the gradually lost ones, the convective motions become weaker after $t=5000$\,s. We collected information about the spicules to analyze their characteristics only during $t=3000-4500$\,s in all the cases.

\subsection{Data processing}

\begin{figure}[H]
   \centering
   \includegraphics[width=9cm, angle=0]{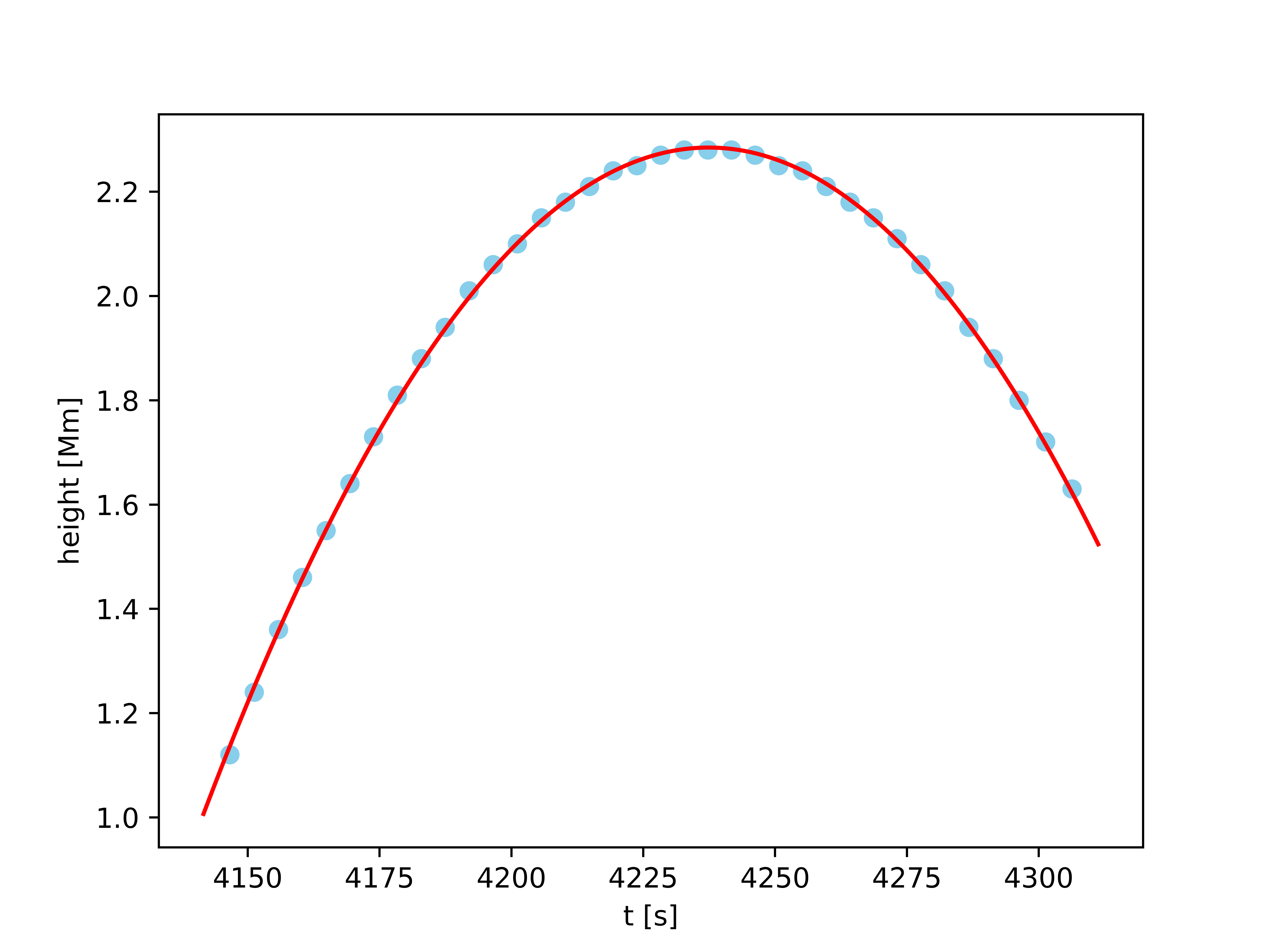}
   \caption{Time-height trajectory of a spicule in Case B. The blue dots represent the height of this spicule at different time points, and the solid red line is the parabolic curve fitted to these data points.}
   \label{fig2}
   \end{figure}

The lifetime of the spicules was determined by manually marking the start and end points of it. We defined the start point of a spicule's lifetime as the first time when its height was greater than or equal to $1$\,Mm and it was not completely obscured by other spicules; the end point of their lifetime was defined as the last time when the height of a spicule was greater than or equal to $1$\,Mm and it was still not yet completely obscured by other spicules. As for the deceleration, maximumu upward velocity and maximum height, we applied the isotherm contour of $lgT=5.0$ as the upper boundary of the spicules. As shown in Figure~\ref{fig2}, the time-dependent height trajectory of each spicule almost follows a parabolic pattern. The maximum height of the spicules was read directly from the trajectories. We performed parabolic fitting on the time-height trajectories, and calculated the deceleration and maximum upward velocity of the spicules based on the fitting results. The time-height trajectories exhibit parabolic characteristics, which means the upward velocity at the top of the spicules decreases as time increases. Therefore, we took the velocities at the start points of their lifetimes as the maximum upward velocities. Additionally, spicules with extremely poor parabolic fitting results were excluded from the statistics in this work.

\section{Numerical Results}

\subsection{The impact of resolution on simulation results}

\begin{figure}[H]
   \centering
   \includegraphics[width=12cm, angle=0]{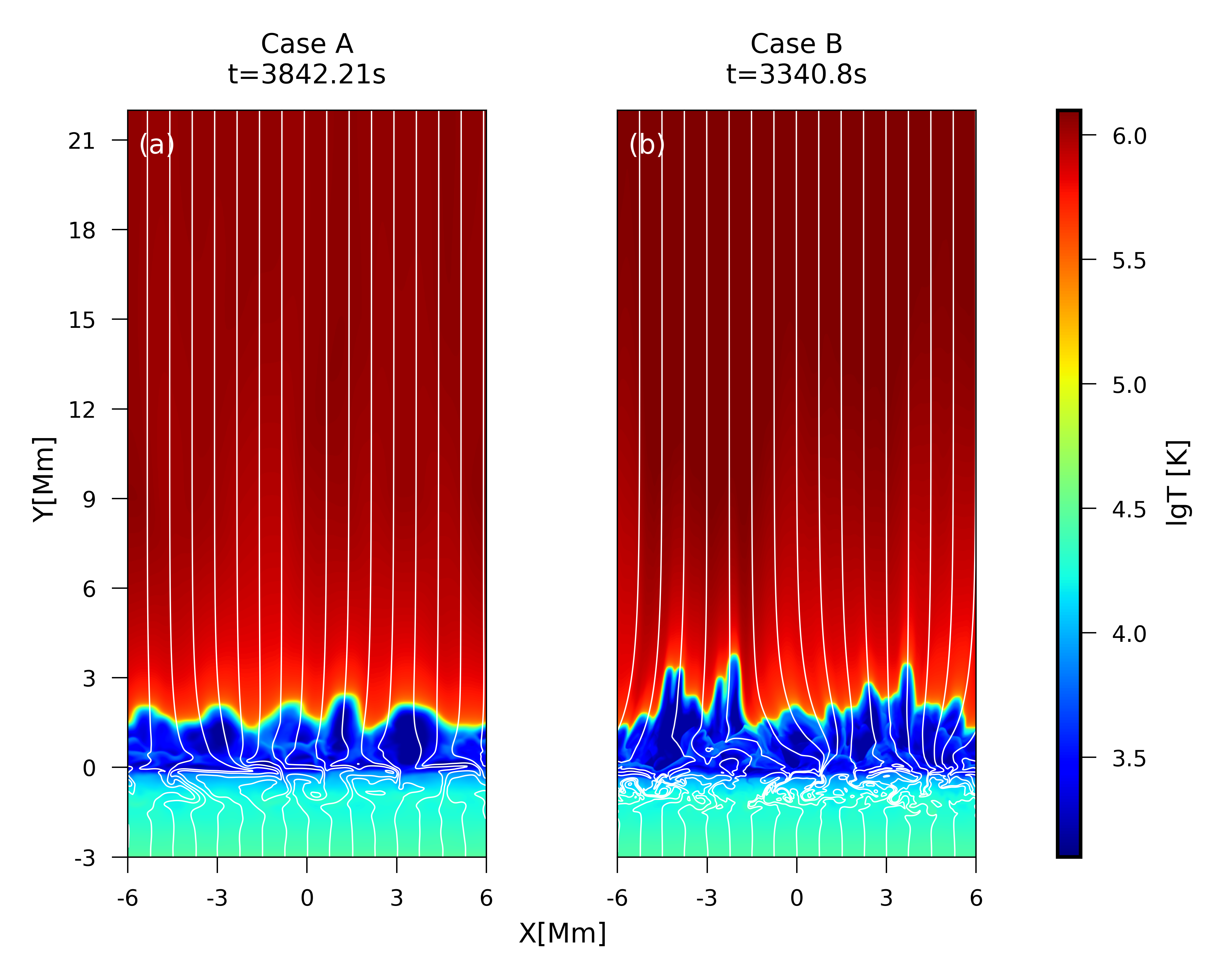}
   \caption{Distributions of the logarithm of temperature in Cases A (left) and B (right) when most of the spicules in the simulation domain are almost at their maximum heights. The white solid lines represent magnetic field lines.}
   \label{fig3}
   \end{figure}

Comparing the results in Cases A and B, we have investigated how resolution settings influence the simulations. Case B has richer details, as shown in Figure~\ref{fig3}, and the spicules in Case B are relatively thinner ($300-500$\,km) and taller. In Case A, there are two or more spicules that are very close to each other, due to the low resolution, appearing as a single spicule, with a width exceeding $1$\,Mm. Observational results show that the widths of solar spicules are mostly concentrated in the range of $200-500$\,km \citep{Pereira2012Quantifying,kuridze2015dynamics}. Therefore, the high-resolution numerical simulation results in Cases B, C, D and E are more consistent with the observational ones. In all the simulation cases, the coronal temperature is maintained at the order of million Kelvin. Our recent numerical simulations \citep{Ni2026} reveal that quasi-periodic spicule upflows trigger formations of slow-mode and shock waves in the corona. These waves are subsequently dissipated via thermal conduction and compression, heating the corona and maintaining its high temperature. Comparing the results from Case A and B, we notice that the average corona temperature is higher in Case B with higher resolution. The reason is that the higher resolution leads to stronger and finer wave structures, which can provide more energy to heat corona.

We should also mention one important point here. The resolution ($\sim12$\,km) in these cases and in the previous high-resolution simulations (e.g., \citealp{martinez-sykora2017generation}) is already very high. However, the resulting numerical diffusivity in all current simulations is still several orders of magnitude greater than the physical magnetic diffusivity calculated from equations (7)-(10) in Section 2.1. Therefore, the magnetic Reynolds number is actually only on the order of $10^4$ in both our simulations and many of the previous simulations with the similar resolution and simulation scale. Future simulations with much higher resolution will likely be required to resolve the physical processes at finer scales.

\subsection{The impacts of background magnetic field on the spicules}

\begin{figure}[H]
   \centering
   \includegraphics[width=13cm, angle=0]{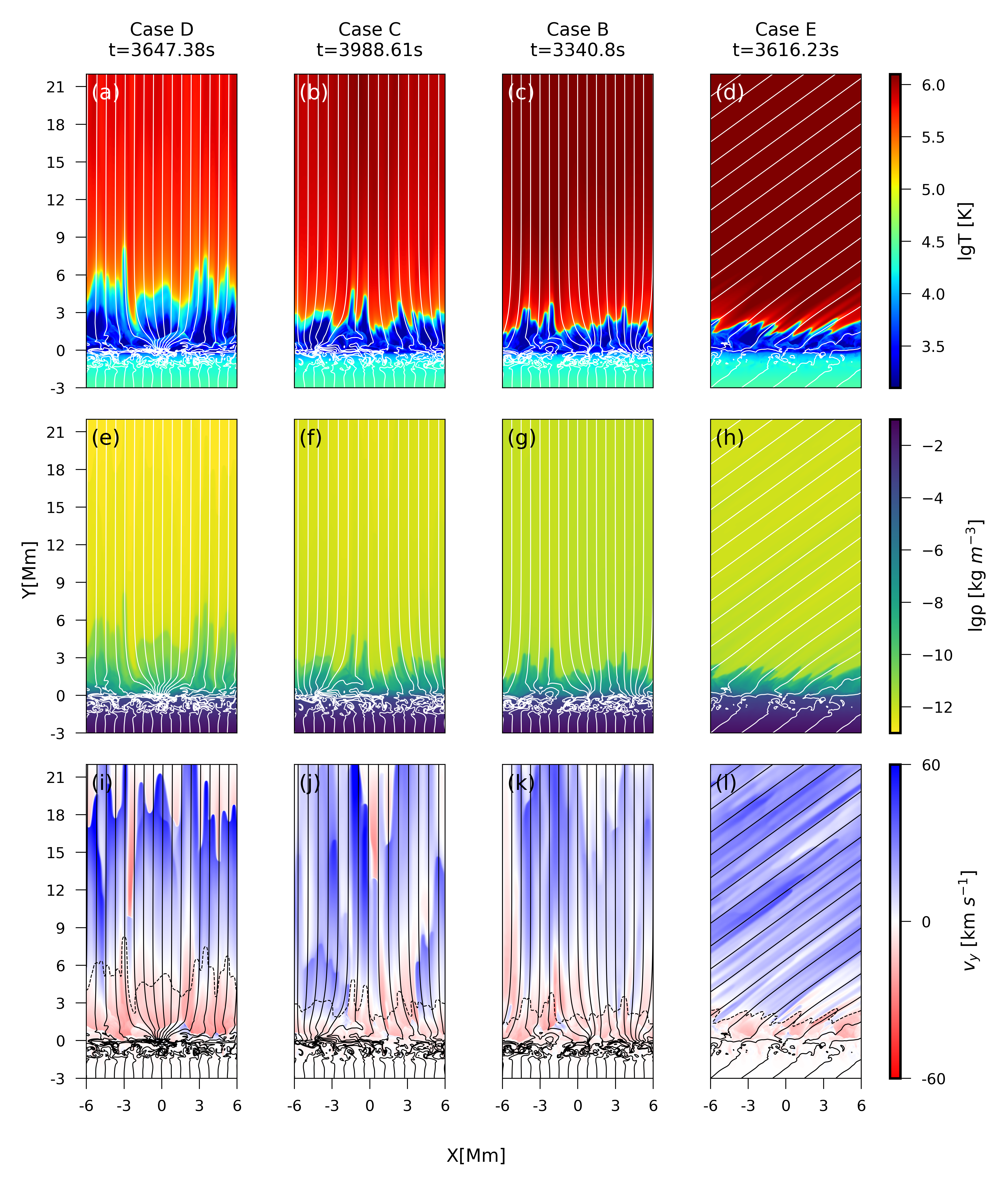}
   \caption{Distributions of the logarithmic of temperature (a-d), the logarithmic of density (e-h) and the velocity in the $y$-direction (i-l) in Cases D, C, B and E. In each panel of the presented figure, most of the spicules in each case are roughly at their maximum height. The white or black solid lines in each panel represent magnetic field lines. The black dashed lines in the bottom panels represent the isotherms where $lgT=5.0$, which denote the upper boundaries of the spicules.}
   \label{fig4}
   \end{figure}

By comparing the results of Cases B, C, D and E, we have investigated how magnetic field strength and inclination influence the properties. Figure~\ref{fig4} shows the logarithmic distributions of temperature, density and the normal velocity in the $y$-direction when most of the spicules are almost at their maximum height in Cases D, C, B and E, respectively. One can observe that the spicules in Case D are longer than those in the other cases. The plasma temperature and density in the corona in Case D are lower than those in the other cases. Therefore, the length of the generated spicules increase with the decreased background magnetic fields. In contrast, the plasma temperature and density in the corona decrease as background magnetic fields weaken. Such results are consistent with the situations measured in the solar atmosphere. As we know, the magnetic fields in active regions are generally stronger than those in quiet Sun regions and coronal holes, where the plasma temperature and density in the corona are relatively lower. The observed spicules in the corona holes and quiet Sun regions are indeed generally longer than those in active regions. The black dashed lines in panels (i), (j), (k), and (l) of Figure~\ref{fig4} represent the isotherms where $lgT=5.0$, which denotes the upper boundaries of the spicules. When these low temperature spicules reach their maximum height, they begin to move downward, while the high-temperature corona plasma above them continues to move upward. 

In Case D with a weaker magnetic field, the hot plasma have higher upward velocities, and some of which exceed $60$\,km\,s$^{-1}$. The background magnetic fields in the corona in Case E (Figure~\ref{fig4}(d), (h), (l)) have an inclination angle of about $53.13^{\circ}$, and the generated spicules exhibit a preferential alignment with the local magnetic field. Comparing the results of Cases B and E, we can find that the spicule heights in the two cases are approximately the same, which means the inclined spicules in Case E are relatively longer. Since the magnetic field strengths in Cases B and E are about the same, the longer spicules in Case E obviously relates to the inclination of the magnetic field.

\begin{figure}[H]
   \centering
   \includegraphics[width=13cm, angle=0]{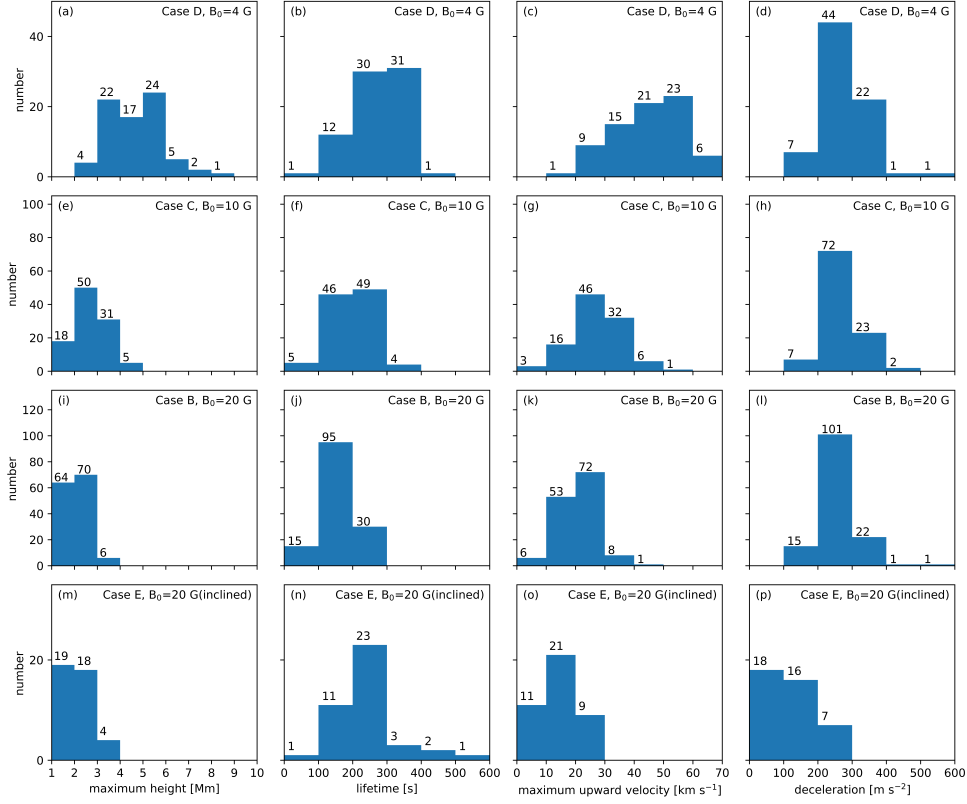}
   \caption{Distributions of the maximum height, lifetime, maximum upward velocity and deceleration of the spicules in Cases D (a-d), C (e-h), B (i-l) and E (m-p).}
   \label{fig5}
   \end{figure}

To quantitatively investigate the effect of background magnetic field on the formed spicules, we have tracked the spicules generated during $t=3000-4500$\,s in Cases B, C, D and E and obtained their decelerations, lifetimes, maximum heights and maximum upward velocities. The specific statistical results are presented in Figure~\ref{fig5}.

The results in panels (a), (e), and (i) of Figure~\ref{fig5} demonstrate the negative correlation between the strength of background magnetic fields and the maximum height of the formed spicules, which is consistent with the conclusion in \citet{Kesri2024ApJ}. In Case B with $B_{0}=20$\,G, the maximum heights of the spicules are mainly concentrated in the range of $1$ to $3$\,Mm; while in Case D with $B_{0}=4$\,G, the majority of the spicules are higher than $3$\,Mm. Most of the spicules in Case D can be classified as type \Roman{t}-like spicules \citep{DePontieu2007TaleTwoSpicules,Pereira2012Quantifying}, or RBEs (e.g., \citealp{kuridze2015dynamics}). The strength of magnetic fields and the maximum upward velocities also show a significant negative correlation as seen in panels (c), (g), and (k) of Figure~\ref{fig5}. In Case D, the velocity of most of the spicules ranges from $40$ to $60$\,km\,s$^{-1}$, which is similar to the observed faster type \Roman{t} spicules \citep{DePontieu2007TaleTwoSpicules, Pereira2012Quantifying}. We also find that the plasma materials in the upper part of the spicules are always heated to a higher temperature during their falling stage, especially in Case D. However, in order to more fully prove that they are type-II spicules, we need to synthesize different spectral line profiles and images and compare them with observations in future work.

In panels (b), (f), and (j) of Figure~\ref{fig5}, we can find that when the strength of the background magnetic fields increases, the lifetime of the spicules decreases. The lifetimes of most of the type \Roman{t}-like spicules in Case D are generally longer than those in other cases. However, previous observations in the low-temperature chromospheric bands indicated that type \Roman{t} spicules, distributed in the regions with a weaker magnetic field, tend to have a shorter lifetime than type \Roman{o} spicules \citep{Pereira2012Quantifying}. One should note that these observations only captured the ascending phase of the type \Roman{t} spicules, resulting in statistics limited to the lifetimes of their ascending phases \citep{Pereira2012Quantifying}, which leads to a significant underestimation for the lifetime of these type \Roman{t} spicules. Later, the observations from satellites such as IRIS and SDO have shown that type \Roman{t} spicules can exhibit responses in both the transition region and coronal wavebands \citep{Pereira2014ApJL,DePontieu2017ApJL,Antolin2018ApJ}, indicating the top part of these spicules are heated to a higher temperature, at least above $10^5$ K \citep{Tian2014Science}. Observations from the high-temperature wavebands have revealed that the lifetime of some type \Roman{t} spicules can even reach $500-800$\,s \citep{Pereira2014ApJL}. In this paper, we have processed the data obtained from numerical simulations, without restrictions on the observation bands. Our simulations include the complete process of ascending and descending, and the measured values can represent the real lifetime of the spicules, which further confirms that the type \Roman{t} spicules have longer lifetimes than type \Roman{o} spicules.

We have also explored the impact of inclined magnetic fields on the spicule formation by comparing the statistical results in Cases B and E. The heights of most of the spicules in these two cases are very close, so the spicules are generally longer in Case E with inclined magnetic fields. The lifetimes of all the spicules in Case B are less than $300$\,s, while there are more than $10$\% of the spicules in Case E with a lifetime exceeding $300$\,s. The decelerations of most of the spicules in Case E are less than $200$\,m\,s$^{-2}$, which is generally smaller than that in Case B with vertical background magnetic fields. Furthermore, the maximum upward velocity of more than half of the spicules in Case B exceeds $20$\,km\,s$^{-1}$, while only about $20$\% of the spicules in Case E reach such a value. In conclusion, the spicules in an environment with inclined background magnetic field are relatively longer, with smaller decelerations and lower maximum upward velocities.

In Figure~\ref{fig5}, one can note that the total number of the spicules are less in the case with weaker background magnetic fields. One reason is that the lifetime of the spicules in the weaker magnetic field case is generally longer and we have analyzed the data during $t=3000-4500$\,s for all cases.

\begin{figure}[H]
   \centering
   \includegraphics[width=13cm, angle=0]{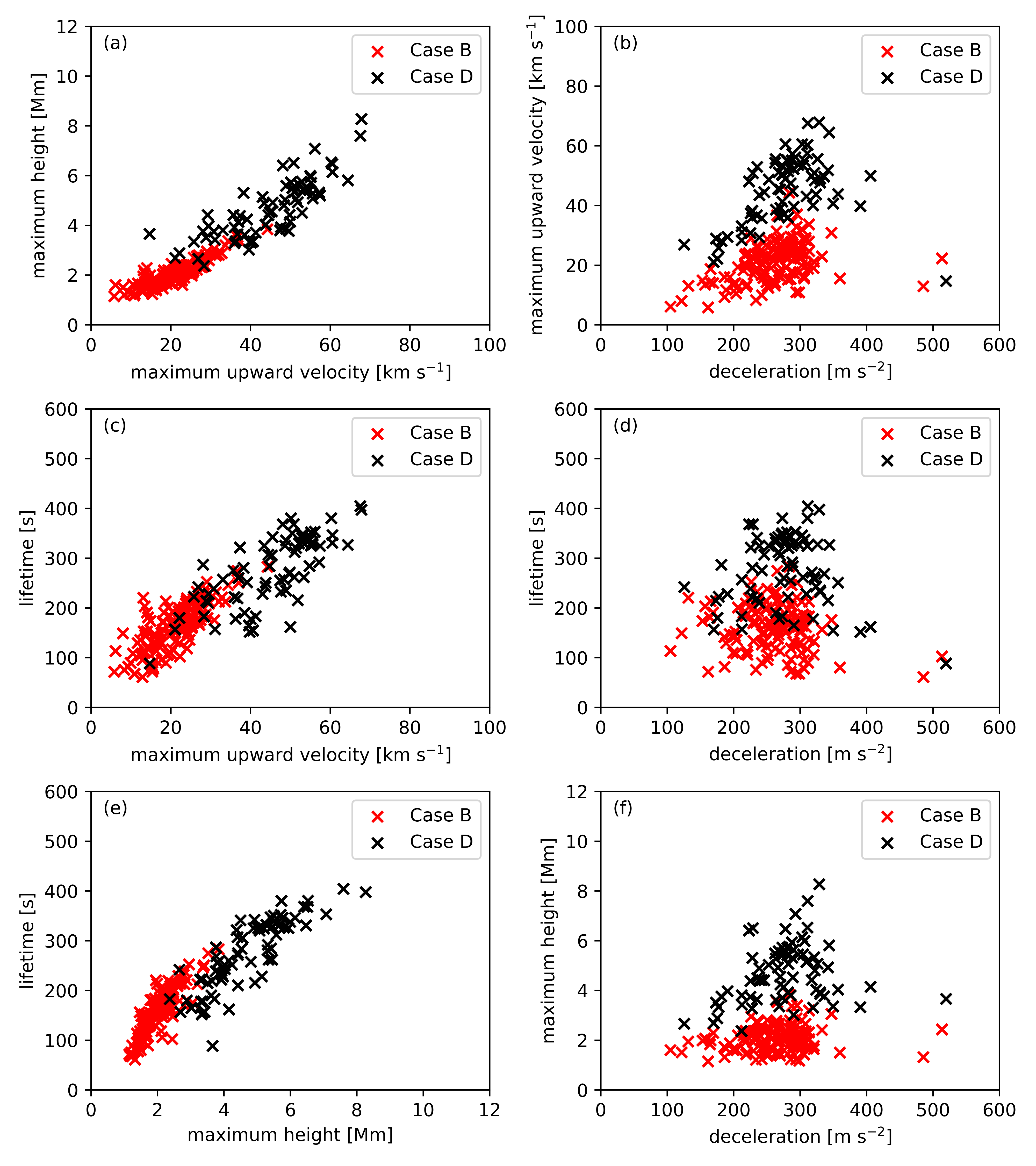}
   \caption{Correlations between any two of the maximum height, lifetime, maximum upward velocity and deceleration of the spicules in Cases B and D.}
   \label{fig6}
   \end{figure}

Figure~\ref{fig6} shows the correlations between any two of the maximum height, lifetime, maximum upward velocity and deceleration of the spicules in Cases B and D. One can find that there are obvious positive correlations between the maximum upward velocity and maximum height, deceleration and maximum upward velocity, maximum upward velocity and lifetime, and maximum height and lifetime (see Figure~\ref{fig6}(a), (b), (c) and (e)), a weak negative correlation between the deceleration and lifetime (see Figure~\ref{fig6}(d)), and no obvious correlation between the deceleration and maximum height (see Figure~\ref{fig6}(f)), which is similar to the results in some previous works (e.g., \citealp{DePontieu2007ApJ,Heggland2007ApJ,MartinezSykora2009ApJ,Heggland2011ApJ,IijimaYokoyama2015}).

\subsection{The driving mechanisms of the spicules}

\begin{figure}[H]
   \centering
   \includegraphics[width=13cm, angle=0]{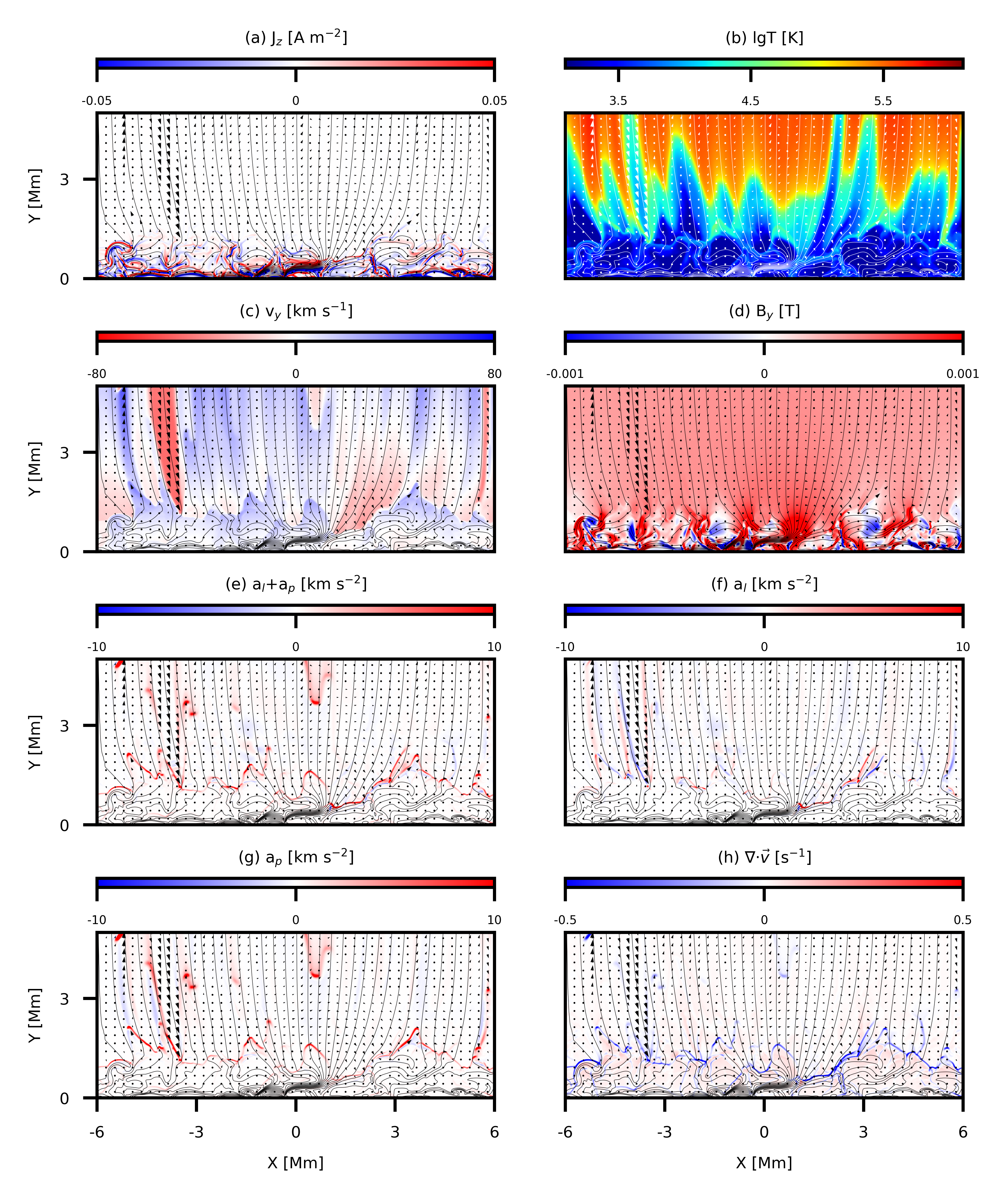}
   \caption{Distributions of different variables when a group of spicules in Case D are in their early rising phase. The current density ($J_z$), the logarithmic of temperature ($lgT$), the velocity in the $y$-direction ($v_y$), the magnetic field in the $y$-direction ($B_y$), the total acceleration in the $y$-direction ($a_l+a_p$), the acceleration in the $y$-direction provided by the Lorentz force ($a_l$), the acceleration in the $y$-direction provided by the pressure gradient ($a_p$) and the divergence of velocity ($\nabla \cdot \boldsymbol{v}$) at $t=3187.22$\,s are presented in panels (a)-(h) respectively. The white or black arrows represent the velocity vector, and the black or white solid lines represent magnetic field lines.}
   \label{fig7}
   \end{figure}

Based on observations and numerical simulations, magnetohydrodynamic shock wave, magnetic reconnection, the upward Lorentz force caused by the shearing and rotational motion of magnetic flux tubes and the Lorentz force amplified by ambipolar diffusion are all considered to be able to drive the generation of spicules. However, it is still not clear which mechanism is dominant in different regions of the Sun.The magnetic field and plasma distribution environment in our numerical simulations are close to that of the solar coronal hole region. By tracking and analyzing the characteristics of each spicule, we can provide theoretical support for investigating the driving mechanisms of the spicules in coronal hole regions.

Figure~\ref{fig7} shows distributions of different variables when a group of spicules in Case D are in their early rising phase. The zoomed-in domain from $0$ to $5$\,Mm in the $y$-direction is presented. At this {moment}, such a group of ascending spicules are at a height of approximately $1.5$\,Mm (see Figure~\ref{fig7}(c)). Figure~\ref{fig7}(h) shows the distribution of the velocity divergence. The value in the blue region is negative, indicating that the plasma here is compressed. Comparing the results in Figure~\ref{fig7}(b), (c) and (h), one can see that many blue regions at around 1.5 Mm in panel (h) basically overlap with the external outlines of the spicules, which demonstrates that the spicule upflows compress the plasma above them when they are rising. Meanwhile, the velocity divergence inside the rising spicule is positive (red region) as shown in panel (h), which indicates that the plasma there is expanding. Therefore, when a spicule is rising, the plasma above it is being compressed, and the plasma beneath the compression zone inside the rising spicule is expanding.

In addition, we have explored the evolutions of different variables below these rising spicules, which can help us to understand their formation process. Panels (f) and (g) of Figure~\ref{fig7} show the distributions of acceleration provided by the Lorentz force and the plasma pressure gradient, respectively. Here, both the Lorentz force and plasma pressure gradient can contribute to the upward movement of the spicules, but in most regions, the contribution of plasma pressure gradient is greater. We also notice that the Lorentz force at some locations are downward which will result in the suppress of the upward acceleration. Since the distributions of magnetic field is complicated in the lower atmosphere, it is not surprise to have the downward Lorentz force at some positions. One can also see such phenomenon in many previous simulations (e.g., \citealp{Ni2026, Iijima2017ApJ}).

Figure~\ref{fig7}(a) presents the current density distribution ($J_z$), one can find that plenty of slender current sheet fragments are generated between $0$ and $1.5$\,Mm in the $y$-direction. The phenomena involving the approach of positive and negative magnetic fields occur at many locations, which exactly correspond to those with strong current density, as shown in panel (d) of Figure~\ref{fig7}. These results indicate that many small-scale magnetic reconnection events may form in this region. The stronger local magnetic field will result in a reconnection process with a higher current density and stronger plasma heating. Magnetic reconnection events do rapidly change the magnetic field topology in surrounding regions. A strong local Lorentz force might be generated to provide an upward acceleration for the plasma. Meanwhile, the high-speed reconnection upward outflows may also create a strong local plasma pressure gradient, lifting the plasma to move upward. Therefore, magnetic reconnection is one possible mechanism to trigger spicules in our simulations.

In addition, the complex convective and turbulent motions around the solar surface can also lead to locally enhanced upward plasma pressure gradients and Lorentz forces. These forces squeeze the plasma to form shock waves. The blue regions with a large value of negative velocity divergence also appear within the range of $0-1$\,Mm in the $y$-direction, as shown in Figure~\ref{fig7}(h), indicating the generation of slow mode shocks there. The Mach number $M_a$ ($M_a=V / V_s$, where $V$ is the absolute value of the total plasma velocity and $V_s$ is the sound speed) in most areas inside this region is larger than $1$. Therefore, the shock compression is another possible mechanism to drive the formation of spicules. 

We have randomly selected $20$ spicules from Case B and another $20$ from Case D respectively, and analyzed the driving mechanisms of each spicule one by one by tracking their formation process. After tracking and analyzing the evolutions of the different variables behind each spicule, we can find that the generation process of most of the spicules relates to both the shock compression and magnetic reconnection. However, one of these two mechanisms may dominate the other, during each spicule formation process. 

\begin{figure}[H]
   \centering
   \includegraphics[width=13cm, angle=0]{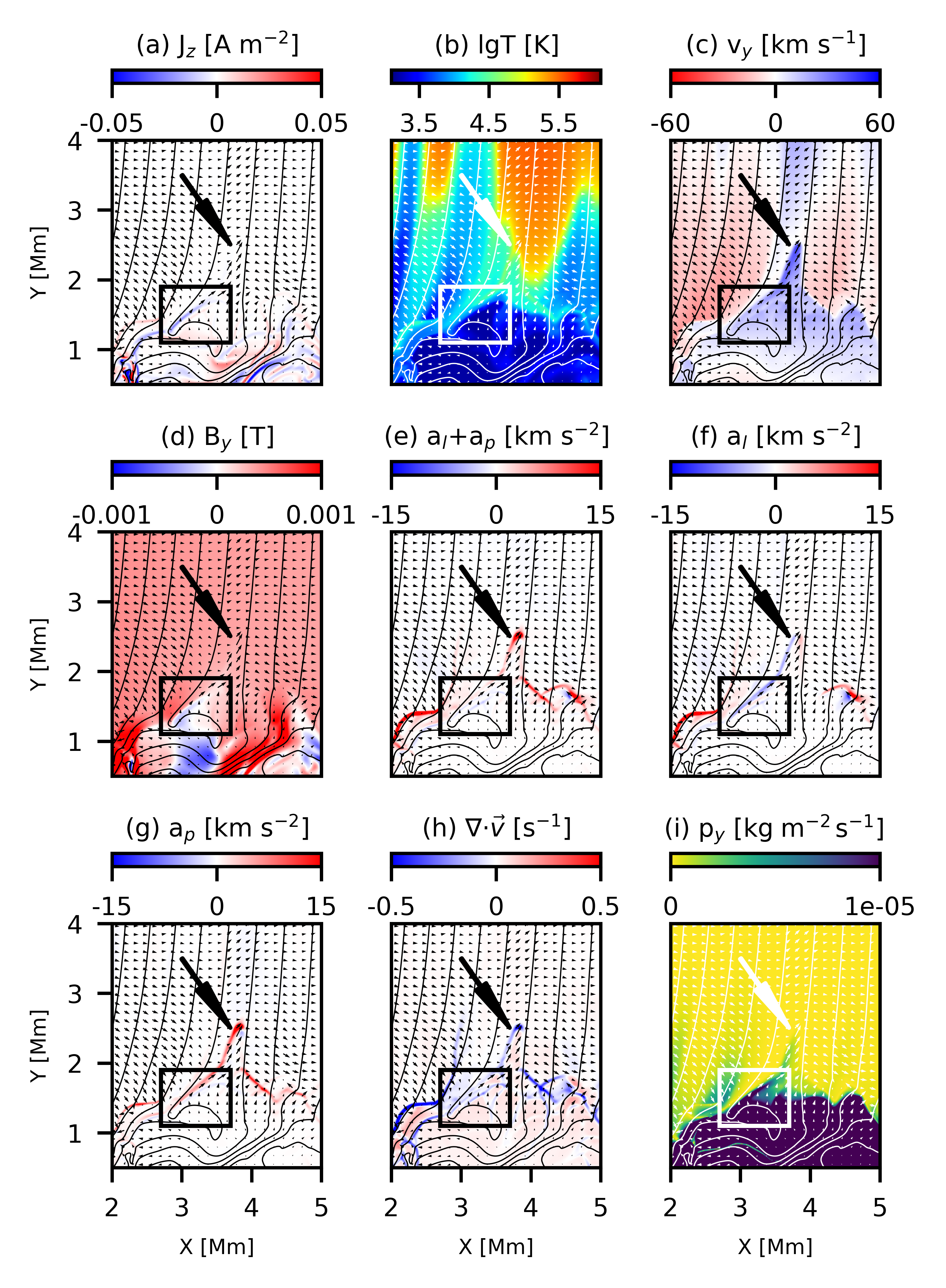}
   \caption{Distributions of different variables at a moment when a spicule in Case D is rising. The current density ($J_z$), the logarithmic of temperature ($lgT$), the velocity in the $y$-direction ($v_y$), the magnetic field in the $y$-direction ($B_y$), the total acceleration in the $y$-direction ($a_l+a_p$), the acceleration in the $y$-direction provided by the Lorentz force ($a_l$), the acceleration in the $y$-direction contributed by the pressure gradient ($a_p$), the divergence of velocity ($\nabla \cdot \boldsymbol{v}$) and the absolute value of momentum density in the $y$-direction ($p_y$) at $t=3187.22$\,s are presented in panels (a)-(i) respectively. The thick big arrow in each panel points to the top of the spicule, and the area within the box is where magnetic reconnection occurs. The white or black arrows represent the velocity vector, and the black or white solid lines represent magnetic field lines. The animation of the corresponding $J_z$ and $p_y$ (Movie8) is available.}
   \label{fig8}
   \end{figure}

In Figure~\ref{fig8}, we show a spicule that is mainly driven by magnetic reconnection in Case D. In each panel of Figure~\ref{fig8}, the magnetic reconnection region that contributes to the formation of this spicule is located inside a small rectangular box, and the thick big arrow points to the top of this spicule. Figure~\ref{fig8}(d) displays the distribution of magnetic field in the $y$-direction, one can see that the positive and negative magnetic fields are close to each other in this region, forming a current sheet as shown in Figure~\ref{fig8}(a). In Figure~\ref{fig8}(g), the strong upward acceleration contributed by plasma pressure gradient is presented on the left side of the spicule. Comparing the results in Figure~\ref{fig8}(e), (f) and (g), we can find that the pressure gradient dominates to result in the upward acceleration for this spicule. Figure~\ref{fig8}(i) and (c) indicate that the high-speed reconnection outflows lead to the formation of the strong plasma pressure gradient and contribute the most part of the selected spicule. The maximum upward velocity of this spicule exceeds $50$\,km\,s$^{-1}$.

\begin{figure}[H]
   \centering
   \includegraphics[width=13cm, angle=0]{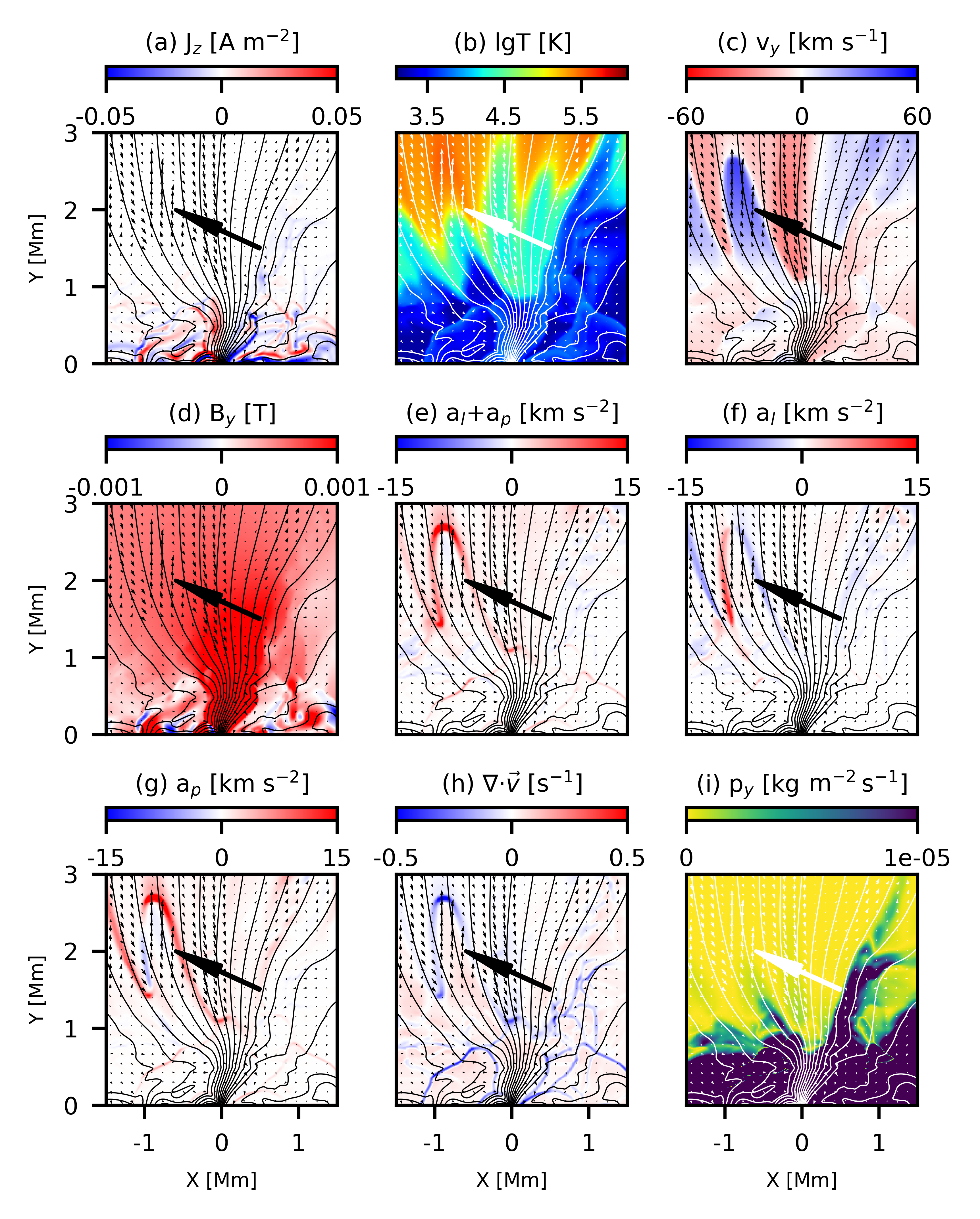}
   \caption{Distributions of different variables at a moment when another spicule in Case D is rising. The current density ($J_z$), the logarithmic of temperature ($lgT$), the velocity in the $y$-direction ($v_y$), the magnetic field in the $y$-direction ($B_y$), the total acceleration in the $y$-direction ($a_l+a_p$), the acceleration in the $y$-direction provided by the Lorentz force ($a_l$), the acceleration in the $y$-direction contributed by the pressure gradient ($a_p$), the divergence of velocity ($\nabla \cdot \boldsymbol{v}$) and the absolute value of momentum density in the $y$-direction ($p_y$) at $t=3119.34$\,s are presented in panels (a)-(i) respectively. The thick big arrow in each panel points to the top of the spicule. The small white or black arrows represent the velocity vector, and the white and black solid lines represent magnetic field lines. The animation of the corresponding $\nabla \cdot \boldsymbol{v}$ and $p_y$ (Movie9) is available.}
   \label{fig9}
   \end{figure}

We show another spicule that is driven mainly by shock waves in Figure~\ref{fig9}, and this spicule is also one of the $20$ spicules randomly selected in Case D. In each panel of Figure~\ref{fig9}, the thick big arrow points to the top of this spicule. Below this spicule, one can find that many small-scale magnetic reconnection events occur near the solar surface (see Figure~\ref{fig9}(a) and (d)). However, as shown in Figure~\ref{fig9}(c) and (i), the upward plasma flows in this spicule do not directly originate from these small scale reconnection outflows. In addition to the significant compression above this spicule, as displayed in Figure~\ref{fig9}(h), we can notice that the blue regions with a large value of negative velocity divergence ($\nabla \cdot \boldsymbol{v}$) also appear below this spicule between $0$ and $1$\,Mm in the $y$-direction, indicating the appearances of slow mode shocks in this region. By tracking the variation of different physical quantities over time and space, we can find that the complex convective and turbulent motions near the solar surface can lead to locally enhanced upward plasma pressure gradients and Lorentz forces, and then trigger the formations of multiple photospheric and chromospheric shock waves. These shock waves can squeeze the plasma above them and finally drive the formation of this spicule. The movie that shows the evolution of the velocity divergence in this region more clearly demonstrates the formation process of this spicule. Additionally, we can find that the maximum upward velocity of this spicule also exceeds $50$\,km\,s$^{-1}$.

\begin{figure}[H]
   \centering
   \includegraphics[width=12cm, angle=0]{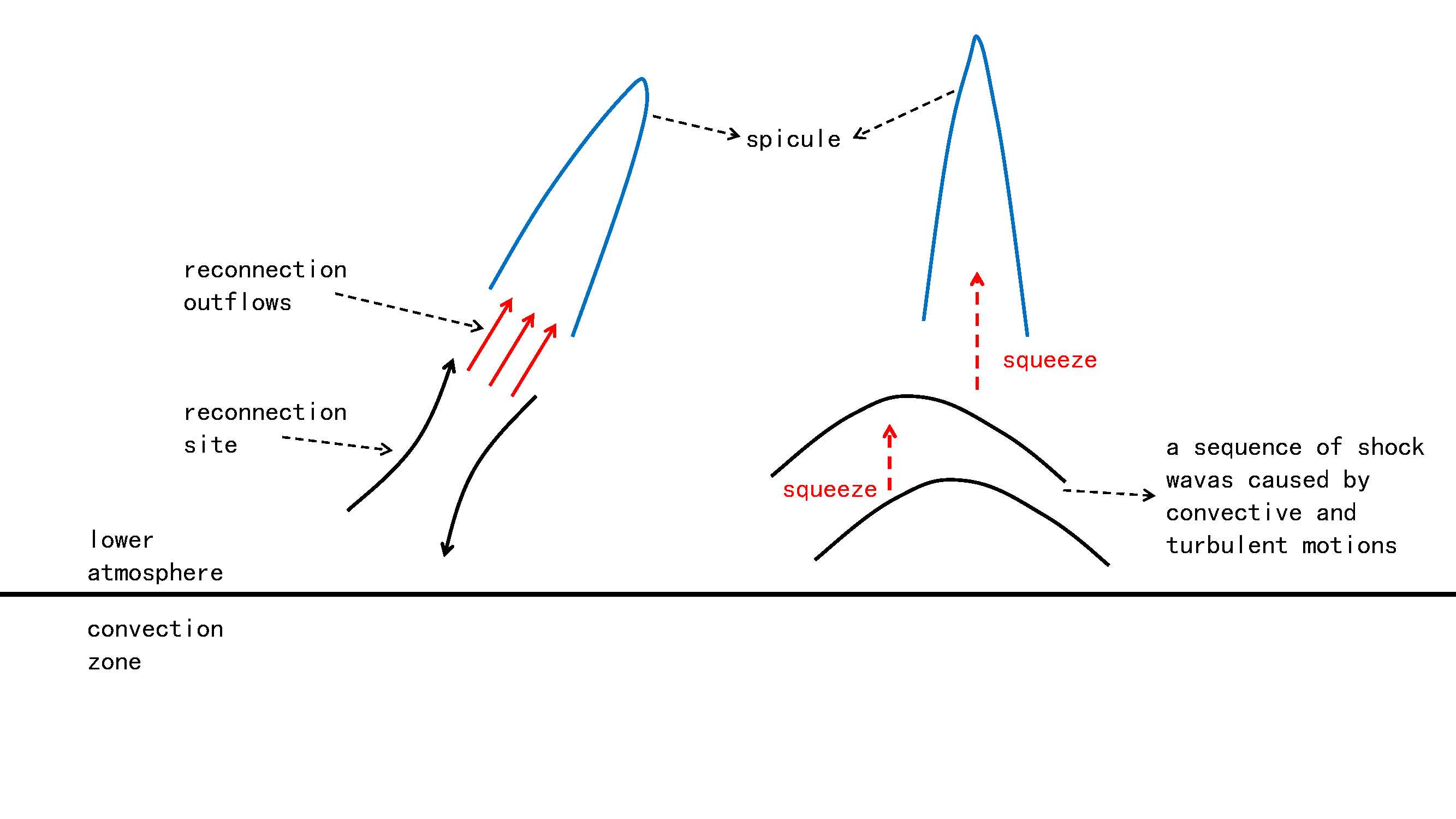}
   \caption{Cartoon schematic of two spicule driving mechanisms. The spicule on the left corresponds to the spicule in Figure~\ref{fig8}, and the spicule on the right corresponds to the one in Figure~\ref{fig9}.}
   \label{fig10}
   \end{figure}

In Figure~\ref{fig10}, we provide a cartoon schematic diagram to more clearly demonstrate the two main driving mechanisms of spicules. We have carefully tracked and analyzed the formation process of each selected spicule. The results show that about one third of the spicules randomly selected in Case D are driven mainly by magnetic reconnection, whereas only about one tenth of the selected spicules in Case B are driven dominantly by magnetic reconnection. More than half of the selected spicules in Cases B and D have strong correlation with shock waves. Therefore, in the coronal hole region where the magnetic field is relatively weak and open, the shock waves triggered by convective and turbulent motions around the solar surface more likely dominate to drive the formations of solar spicules. However, the formation of a small portion of spicules is still mainly contributed by magnetic reconnection outflows.

\section{Summary and discussion}
In this paper, we have studied the formation and properties of solar spicules through 2.5D radiative MHD simulations. The simulation region includes the upper convection zone, photosphere, chromosphere, and lower corona. We employed distinct radiation cooling models for the upper convection zone, photosphere, chromosphere, and lower corona, respectively, to more accurately replicate the real solar atmospheric radiation environment. The convective motions were self consistently generated to drive spicules. We experimented with five cases. By varying the strength and inclination of the initial magnetic field in each case, we have investigated the impact of background magnetic fields on the formed spicules. After the system has reached dynamic equilibrium, we have tracked and derived the characteristics of the spicules such as the maximum height, lifetime, deceleration and the maximum upward velocity. Additionally, we also have tracked the formation process of each selected spicule in two different cases to identify their primary driving mechanisms. Based on the research findings, the main conclusions are as follows:

1. The convective and turbulent motions near the solar surface can cause many local small-scale magnetic reconnection events and shock wave structures in the photosphere and chromosphere. In the coronal hole region, where the background magnetic field of the corona is open, shock waves dominate to drive the formation of the spicules, while less than one third of the spicules are driven mainly by reconnection outflows.

2. In environments with weaker background magnetic fields, the spicules exhibit higher heights, longer lifetimes, and higher maximum rising velocities. These results are consistent with the solar observations, showing that the longer and faster spicules normally appear in corona holes and quiet Sun regions with weaker magnetic fields.

3. When the background magnetic field is inclined, the generated spicules incline accordingly, and they are relatively longer, with smaller decelerations and lower maximum upward velocities.

4. There are obvious positive correlations between the maximum upward velocity and maximum height, deceleration and maximum upward velocity, maximum upward velocity and lifetime, and maximum height and lifetime of the spicules, a weak negative correlation between the deceleration and lifetime, and no obvious correlation between the deceleration and maximum height.

The statistical characteristics of solar spicules and how they vary with the background magnetic field strength and inclination from our simulations are consistent with previous observational and numerical results (e.g., \citealp{Pereira2012Quantifying,Heggland2011ApJ,Kesri2024ApJ}).

Although for driving solar spicules several different physical mechanisms have been proposed, such as MHD shock wave, magnetic reconnection, the upward Lorentz force caused by the shearing and rotational motion of magnetic flux tubes and the Lorentz force amplified by ambipolar diffusion (see e.g., \citealp{Hollweg1982Origin,DePontieu2004Nature,Singh2011PhPl,martinez-sykora2017generation,Iijima2017ApJ,Liu2019Evidence,Shen2021RSPA,Samanta2019Science,Oxley2020ApJ,Scalisi2021ApJ,Dey2022NatPhys}), the dominant mechanism remains unclear. The complete physical picture of how spicules are generated in a real complex solar environment has not yet been fully understood. 

The formation process of spicules is closely related to convective and turbulent motions near the solar surface. In this work, we can find that these motions can lead to many local MHD shock waves and small-scale magnetic reconnection events in the lower solar atmosphere, and they normally work together to trigger the formation of spicules. By carefully tracking the formation process of dozens of the spicules, we also can find that, in the coronal-hole-like environments, these frequently occurring shock waves dominate to drive the formation of the spicules, while less than one third of the spicules are driven mainly by reconnection outflows. Such research findings provide a more in-depth theoretical basis for the understanding of the entire formation process of solar spicules.

Though our numerical studies can approximately present the real formation process of solar spicules in the coronal hole region, there is still plenty of room for the improvements of the numerical model and the settings of boundary conditions. We are working on developing the NIRVANA code by including a more suitable tabulated equation of state and partial ionization treatment for the convection zone and photosphere. It is also important to set a more suitable bottom boundary condition by replenishing the gradually lost mass and energy in the simulation domain, then we can get a much longer simulation run with continuous reasonable convective motions. The ambipolar diffusion effect in the partially ionized lower solar atmosphere is not included in this work. We conjecture that fully implementing ambipolar diffusion in the simulations may result in somewhat different spicule lengths, however, this still may not change our main conclusions. Additionally, the convective and turbulent motions may also cause shearing and rotational motions of magnetic flux tubes, which needs to be investigated in future 3D simulations.

\normalem
\begin{acknowledgements}
Lei Ni thanks Dr. Yajie Chen from Max-Planck Institute for the discussions about boundary conditions. This research is supported by the Strategic Priority Research Program of the Chinese Academy of Sciences with Grant No.XDB0560000; the NSFC Grants No.12373060; the National Key R\&D Program of China with Grant No.2022YFF0503804(2022YFF0503800); the National Key R\&D Program of China with Grant No.2022YFF0503003 (2022YFF0503000); the outstanding member of the Youth Innovation Promotion Association CAS (No.Y2021024); the Basic Research of Yunnan Province in China with Grant No.202401AS070044; the Yunling Talent Project for the Youth; the Yunling Scholar Project of the Yunnan Province and the Yunnan Province Scientist Workshop of Solar Physics; Yunnan Key Laboratory of Solar Physics and Space Science under the number 202205AG070009; the China's Space Origins Exploration Program. The simulation work was carried out at National Supercomputer Center in Tianjin, and the calculations were performed on Tianhe new generation supercomputer. The numerical data analysis have been done on the Computational Solar Physics Laboratory of Yunnan Observatories. This research is grateful to NKFIH OTKA (Hungary, grant No.K142987); the Science and Technology Facilities Council (STFC, grant No. ST/M000826/1) UK; PIFI (China, grant number 2024PVA0043) and the NKFIH (Hungary) Excellence Grant (grant nr TKP2021-NKTA-64). We acknowledge support from ISSI-Beijing for the project on "Small-scale eruptions in the Sun" (ID 24-604).
\end{acknowledgements}

\bibliographystyle{raa}
\bibliography{ms2026-0113}

\label{lastpage}

\end{document}